\documentclass[12pt]{article}
\usepackage[utf8]{inputenc}
\usepackage{graphicx}
\usepackage{animate}
\usepackage{amsmath,bm,amsfonts,amssymb,enumerate,epsfig,bbm,calc,color,ifthen,capt-of,multimedia,bm,multirow}
\usepackage{blindtext}
\usepackage{booktabs}
\usepackage{MnSymbol}
\usepackage{url}
\usepackage{natbib}
\usepackage{changepage}
\usepackage{fancyhdr}
\usepackage[flushleft]{threeparttable}

\usepackage{setspace}
\usepackage[flushleft]{threeparttable}

\DeclareMathOperator*{\arginf}{arg\,inf}

\title{Robust estimation of optimal dynamic treatment regimes with nonignorable missing covariates}
\author{Jian Sun \thanks{School of Data Science,  Fudan University, Shanghai 200433, China. Email: jsun19@fudan.edu.cn.}, Bo Fu \thanks{School of Data Science,  Fudan University, Shanghai 200433, China. Email: fu@fudan.edu.cn.}, and Li Su \thanks{MRC Biostatistics Unit, University of Cambridge, Cambridge, CB2 0SR, UK. Email: li.su@mrc-bsu.cam.ac.uk.}}

\date{}

\begin{document}

\maketitle

\begin{abstract}
Estimating optimal dynamic treatment regimes (DTRs) using observational data (e.g., from electronic medical records)  is often challenged by nonignorable missing covariates arsing from informative monitoring of patients in clinical practice.  To address nonignorable missingness of pseudo-outcomes induced by nonignorable missing covariates, a weighted Q-learning approach using parametric Q-function models and a semiparametric missingness propensity model has recently been proposed. However, misspecification of parametric Q-functions at later stages of a DTR  can propagate estimation errors to earlier stages via the pseudo-outcomes themselves and indirectly through biased estimation of the missingness propensity of the pseudo-outcomes. This robustness concern motivates us to develop a direct-search-based optimal DTR estimator built on a robust and efficient value estimator, where nonparametric methods are employed for treatment propensity and Q-function estimation, and inverse probability weighting is applied using missingness propensity estimated with the aid of nonresponse instrumental variables. Specifically, in our value estimator, we replace weights estimated by prediction models of treatment propensity with stable weights estimated by balancing covariate functions in a reproducing-kernel Hilbert space (RKHS). Augmented by Q-functions estimated by RKHS-based smoothing splines, our value estimator mitigates the misspecification risk of the weighted Q-learning approach while maintaining the efficiency gain from employing pseudo-outcomes in missing data scenarios. The asymptotic properties of the proposed estimator are derived, and simulations demonstrate its superior performance over weighted Q-learning under model misspecification.  Simpler versions of the proposed estimator developed for optimal single-stage treatment rule estimation with nonignorable missing covariates also perform better than competitor estimators in simulations. We apply the proposed methods to investigate the optimal fluid strategy for sepsis patients using data from the Medical Information Mart for Intensive Care database. 
\end{abstract}


\section{Introduction} \label{Sec_1}


\subsection{Nonignorable missing covariates in optimal dynamic treatment regime estimation} \label{motivatingdata}
Sepsis is a life-threatening syndrome caused by the body’s response to infection, resulting in damage to its own tissues and organs. Timely and effective fluid resuscitation is crucial for stabilizing tissue hypoperfusion/shock (e.g., inadequate delivery of vital oxygen and nutrients to body tissues)   in sepsis patients. The `Surviving Sepsis Campaign' guideline strongly recommends administrating at least 30 mL/kg of intravenous (IV) fluid within the first 3 hours of intensive care unit (ICU) admission \citep{Rhodes_et_al_2017}. However, the most beneficial fluid resuscitation strategy in the early hours of treatment of sepsis patients remains unknown.
To address this lack of evidence, \cite{Speth_et_al_2022} analyzed electronic medical record data from the Medical Information Mart for Intensive Care (MIMIC-III) \citep{Johnson_et_al_2016} to estimate an optimal two-stage dynamic treatment regime (DTR) of fluid resuscitation in adult septic patients  during the 0-3 hours and 3-24 hours following admission to the medical ICU. However, hemodynamic variables (e.g., blood pressure, respiratory rate and body temperature), which should guide additional fluid administration after initial resuscitation as recommended by the `Surviving Sepsis Campaign', were left out in the optimal DTR by  \cite{Speth_et_al_2022}. Moreover, certain hemodynamic variables in the MIMIC-III data were partially missing at the conclusion of the initial resuscitation, and their missingness was likely to be nonignorable because of the informative monitoring of the patients in the MIMIC-III database. Specifically, patients with normal physiological indicators and less severe conditions received less intensive monitoring. Hence, their measurements of hemodynamic variables were more likely to be missing, and this missingness was likely nonignorable because it was directly related to the unmeasured values of hemodynamic variables.

Motivated by the challenge of incorporating hemodynamic variables  to improve the estimation of optimal fluid resuscitation strategy using the MIMIC-III data, \cite{sun2025weighted} developed a weighted Q-learning approach, which, to the best of our knowledge,  was the first attempt to handle nonignorable missing covariates for optimal DTR estimation.  \cite{sun2025weighted} pointed out that, under the future-independent missingness assumption (i.e., covariate missingness does not depend on variables measured in the future),  a complete-case analysis is still valid for estimating an optimal single-stage rule or the final-stage  rule of a DTR.  However,  the pseudo-outcomes at earlier stages of Q-learning are also subject to nonignorable missingness because these pseudo-outcomes from backward-induction are functions of the nonignorable missing covariates at later stages. Therefore,  to address the nonignorable missing pseudo-outcomes,  \cite{sun2025weighted} proposed a semiparametric missingness propensity model and estimated the missingness propensity of the pseudo-outcomes with the aid of nonresponse instrumental variables.  In both simulations and the analysis of the MIMIC-III data, their weighted  Q-learning approach using the inverse missingness propensity demonstrated better performance with larger expected final outcomes under the estimated DTR than other Q-learning approaches based on a complete-case analysis and multiple imputation.  

In the presence of missing covariates, an important benefit of  Q-learning methods with pseudo-outcomes is that the estimation of the decision rule at stage $t$ is affected by the covariate missingness at stage $t+1$, but not by that from stage $t+2$ to the final stage. As a result, there could be more available data for analysis in Q-learning, compared to other DTR methods not involving pseudo-outcomes. In contrast,   DTR methods relying on observed outcome data, such as simultaneous outcome weighted learning \citep{zhao2015new}, would only be able to include patients with complete covariate data \textit{throughout all stages}. This would exacerbate the sample size concern when there might be only a small proportion of patients who would follow a specific DTR in observed data.    

Nevertheless,  the weighted Q-learning approach by \cite{sun2025weighted} 
 was based on parametric Q-function models, which are vulnerable to model misspecification and raise robustness concerns for practical use. First,   Q-learning and other regression-based methods such as A-learning share a problem: the form of the estimated outcome-regression function directly dictates the form of the estimated optimal treatment regime. Hence researchers face a trade-off in regression modeling---a parsimonious model risks misspecification, whereas a complex model may yield unintelligible treatment regimes \citep{luckett2020estimating}. In contrast, direct-search methods decouple the form of decision rules from the estimation of treatment value functions \citep{zhang2013robust,zhao2015new,liu2018augmented,zhang2018c}, which may offer more robustness to model misspecification than Q-learning.  
Second,  Q-function misspecification at later stages propagates estimation errors to earlier stages through pseudo-outcomes. These errors not only directly affect Q-function estimation at earlier stages but also indirectly do so via influencing the missingness propensity estimation. Consequently, with nonignorable missing covariates,  misspecification in later-stage Q-functions may amplify the inconsistency of earlier-stage decision rule estimation.

\subsection{Overview of contributions}
Motivated by the robustness concerns of the weighted Q-learning approach by \cite{sun2025weighted}, we propose direct-search methods of optimal single-stage and multi-stage treatment rules with nonignorable missing covariate: nonparametric methods for nuisance parameter estimation are employed to construct estimators of treatment regime values, and then optimal decision rules are subsequently estimated by a classification-based estimator. Specifically, under the future-independent missingness assumption and using the complete-case sample, we propose two estimators of the optimal single-stage rule, Covariate-Functional-Balancing Learning (CFBL) and Augmented Covariate-Functional-Balancing Learning (ACFBL),  where 
treatment regime values are estimated by a balancing weights estimator and an augmented balancing weights estimator, respectively. The \textit{balancing weights} optimize balance between treatment groups over a
reproducing-kernel Hilbert space (RKHS) of covariate functions spanning the space of potential outcome means, which are more stable and robust than weights estimated by prediction models for treatment propensity \citep{wong2018kernel}.  For average treatment effect (ATE) estimation, balancing weights are empirically and theoretically shown to be more stable than weights obtained by maximum likelihood estimation in scenarios with model misspecification and/or strong confounding, which considerably improves the performance of inverse probability weighted estimators of ATE   (e.g., see \citealp{imai2014covariate, chan2016globally, Yiu2017,  Chattopadhyay2020,Tan2017}).
The simulation study in Section~\ref{simulation} also demonstrates that our CFBL and ACFBL estimators outperformed Outcome-Weighted Learning (OWL) of \cite{zhao2015new} and Efficient Augmentation and Relaxation Learning (EARL) of \cite{zhao2019efficient} based on value estimators with inverse treatment propensity weights estimated by support vector machines. The proposed ACFBL estimator is augmented by Q-function estimates from RKHS-based smoothing spline methods, which mitigates the misspecification concern about parametric Q-function models.

For optimal multi-stage rules, we  propose an ACFBL estimator to benefit from the sample size and efficiency gain of utilizing pseudo-outcomes in missing data 
scenarios. To address nonignorable missingness of covariates and pseudo-outcomes, we follow \cite{sun2025weighted} and employ inverse probability weighting with weights estimated by a semiparametric missingness propensity model and nonresponse instrumental variables.  By incorporating RKHS-based smoothing spline methods for Q-function, our ACFBL estimator reduces the risk of model misspecification propagating to earlier stages through pseudo-outcomes themselves and biased estimates of the missingness propensity.  Coupled with the balancing weights, this makes our ACFBL estimator a robust alternative to the weighted Q-learning approach by \cite{sun2025weighted}.

\subsection{Related methods}

In the literature, balancing weights were adopted by \cite{wang2023projected} to tackle the expanded dimension issue in infinite-stage offline reinforcement learning under the Markov assumption. 
 \cite{chu2023targeted} and \cite{chen2024robust}  developed balancing weights to generalize optimal single-stage rules estimated in a source population to a target population with covariate shifts. None of these methods is concerned about missing data issues in optimal treatment rule estimation. 

With ignorable missing data, multiple imputation \citep{Shortreed_et_al_2014,Shen_Hubbard_Linn_2023} and augmented inverse probability weighting \citep{Dong_et_al_2020,Huang_Zhou_2020} approaches have been applied for estimating optimal DTRs. 
However, these methods could potentially yield sub-optimal DTRs when missing data for covariates, treatments or outcomes are nonignorable  \citep{Little_Rubin_2014, sun2025weighted}. For handling missingness in covariates, treatments and outcomes in sequential multiple assignment randomized trials, \cite{Shortreed_et_al_2014} proposed a time-ordered nested conditional imputation strategy when estimating optimal DTRs. \cite{Dong_et_al_2020} applied the augmented inverse probability weighting approach to Q-learning and a generalized version of outcome-weighted learning when dealing with missing data caused by patients' dropout. For single-stage scenarios with missing data, 
\cite{Shen_Hubbard_Linn_2023} extended the multiple imputation method proposed by \cite{Shortreed_et_al_2014}  to estimate optimal treatment rules that were not directly observed in the design. \cite{Huang_Zhou_2020} investigated the performance of an augmented inverse probability weighted estimator in the direct-optimization framework for optimal single-rule estimation with missing covariates.  None of these methods is applicable to handle the nonignorable missing hemodynamic variables for estimating optimal  fluid resuscitation strategy for adult sepsis patients in the MIMIC-III data.

\section{Methods}

\subsection{Setting and notation}


We consider an observational cohort comprising $n$ patients undergoing a finite sequence of $T$ treatment stages. The subscripts $i = 1, \ldots, n$ and $t = 1, \ldots, T$  are used to index individual patients and treatment stages, respectively. Following standard notation, we employ capital letters to denote random variables and lowercase letters to represent specific realizations of these variables. 
For each patient, let $(\bm{X}_t,A_t,Y_t)$ be the complete data measurable at stage $t$ for $t = 1, \ldots, T$, where $A_t\in \{-1, 1\}$ is the assigned treatment,  $\bm{X}_t \in \mathbb{R}^{p_t}$ is a  
$p_t$-dimensional 
vector of covariates measured before $A_t$, and $Y_t$ denotes the longitudinal outcome measured after $A_t$. 
A summary of notations is provided in Table S1 of the Supplementary Materials. 

The final outcome of interest is a pre-specified function (e.g., sum or maximum) of the longitudinal outcomes in $T$ stages, $Y = f(Y_1,\ldots, Y_T)$. Larger values of $Y$ are assumed to be better. We define $\bm{H}_1=\bm{X}_1$, and $\bm{H}_t=\left(\bm{H}_{t-1}, A_{t-1}, Y_{t-1}, \bm{X}_t\right)$ for $t=2, \ldots, T$. Thus, $\bm{H}_t$ represents the information available before making the treatment decision at stage $t$. Let $\mathcal{H}_t$ denote the support of $\bm{H}_t$. A DTR $\bm{d}$ consists of a set of decision rules $(d_1, \ldots, d_T)$, where $d_t: \mathcal{H}_t \to \{-1, 1\}$ is a function that takes the observed history $\bm{h}_t$ as input and outputs a treatment decision at stage $t$. An optimal DTR is the set of decision rules $\left(d^{\mathrm{opt}}_1, \ldots, d^{\mathrm{opt}}_T\right)$ that maximizes the expectation of the final outcome. We formalize this definition using the potential outcome framework.

For $t=1,\ldots,T$, let $Y_t^*(\bar{\bm{a}}_t)$  denote the potential  outcome at stage $t$ and $\bm{X}_{t+1}^*(\bar{\bm{a}}_t)$  denote the potential covariates at stage $t+1$  if a patient, possibly contrary to fact, had received the treatment sequence $\bar{\bm{a}}_t = (a_1,\ldots,a_t)$  by stage $t$. The set of potential outcomes under $\bar{\bm{a}}_t$ is $\bm{O}_t^*(\bar{\bm{a}}_t) = \{Y_1^*(a_1), \bm{X}_2^*(a_1),\ldots,Y_{t}^*(\bar{\bm{a}}_{t}),\bm{X}_{t+1}^*(\bar{\bm{a}}_{t}) \}$  for $t=1,\ldots,T-1$. The potential final outcome under a regime $\bm{d}$ is
\begin{align*}
Y^*(\bm{d})=\sum_{\bar{\bm{a}}_T} Y^*\left(\bar{\bm{a}}_T\right) \prod_{t=1}^T \mathbbm{I} (d_t\left[\{\bm{X}_1,a_1,Y_1^*(a_1),\ldots,a_{t-1},Y_{t-1}^*(\bar{\bm{a}}_{t-1}),\bm{X}_{t}^*(\bar{\bm{a}}_{t-1})\}\right]=a_t),
\end{align*}
where $Y^*\left(\bar{\bm{a}}_T\right)=f\left\{Y_1^*\left(a_1\right), Y_2^*\left(\bar{\bm{a}}_2\right), \ldots, Y_T^*\left(\bar{\bm{a}}_T\right)\right\}$ and $\mathbbm{I}(\cdot)$ is an indicator function. 
An optimal DTR, $\bm{d}^{\mathrm{opt}}\in \mathcal{D}$, satisfies that $E\{Y^*(\bm{d}^{\mathrm{opt}})\} \geq E\{Y^*(\bm{d})\}$ for all $\bm{d} \in \mathcal{D}$. 
Note that this definition of optimality depends on the class $\mathcal{D}$. To estimate the value of candidate regimes from observed data, it is necessary to express the value of regimes solely in terms of observables rather than potential outcomes. This becomes feasible under the following, now standard, causal assumptions:
\begin{description}
    \item {\em Assumption} 1 (Consistency). $Y_t^*(\bar{\bm{a}}_t) = Y_t$ for $t = 1,\ldots,T$ and $\bm{X}_{t+1}^*(\bar{\bm{a}}_t) = \bm{X}_{t+1}$ for $t = 1,\ldots,T-1$ when $\bar{\bm{a}}_t$ are actually received.
    \item {\em Assumption} 2 (Sequential ignorability). $\{\bm{O}_{T-1}^*(\bar{\bm{a}}_{T-1}), Y_T^*(\bar{\bm{a}}_T) : \bar{\bm{a}}_T \in \bigotimes_{t=1}^{T} \{-1,1\}\} \perp\!\!\!\perp A_t \mid \bm{H}_t$ for $t=1,\ldots,T$, where $\bigotimes$ denotes the Cartesian product.
    \item {\em Assumption} 3 (Positivity). $P(A_t=a_t\mid \bm{h}_t) > \epsilon >0$ for $a_t \in \{-1,1\}$, $\bm{h}_t \in  \mathcal{H}_t$, and $t=1,\ldots,T$, where $\epsilon$ is a positive constant. 
\end{description}

Under Assumptions 1-3, the value of a DTR can be identified from the observed data. Specifically, in the 2-stage scenario we have $E\{Y^*(\bm{d})\} = E(E[\{Y\mid \bm{H}_2,A_2 = d_2(\bm{H}_2)\}\mid \bm{H}_1,A_1 = d_1(\bm{H}_1)] )$.
This iterative conditional expectation form suggests that the optimal DTR can be computed by a backwards-induction procedure, which is based on the following recursively defined Q-functions. 
At the final stage $T$, $Q_T\left(\bm{h}_T, a_T\right) = E\left(Y \mid \bm{H}_T=\bm{h}_T, A_T=a_T\right)$. For $ t=T-1,\ldots,1$, $Q_t\left(\bm{h}_t, a_t\right) = E\{\max_{a_{t+1} \in \{-1,1\}} Q_{t+1}(\bm{H}_{t+1}, a_{t+1}) \mid \bm{H}_t=\bm{h}_t, A_t=a_t\}.$ The standard Q-learning approach employs regression models to estimate these Q-functions. Note that in these models, the outcome variable is $\max _{a_{t+1} \in \{-1,1\}} Q_{t+1}\left(\bm{H}_{t+1}, a_{t+1}\right)$ instead of $Y_t$ ($t = 1,\ldots,T-1$). These outcome variables are commonly referred to as pseudo-outcomes, and we define $Y^{pse}_T = Y$ and $Y^{pse}_t = \max _{a_{t+1} \in \{-1,1\}} Q_{t+1}\left(\bm{H}_{t+1}, a_{t+1}\right)$ for $t = 1,\ldots,T-1$. Subsequently, $Q_t(\bm{h}_t,a_t) = E\left( Y^{pse}_t \mid \bm{h}_t,a_t \right)$ for $t = 1,\ldots,T$. 


\subsection{Equivalence of the optimal decision rules in the complete-case and full samples} \label{CCsame}

We consider a scenario in which both treatment and longitudinal outcome variables, $(A_t, Y_t)$, are fully observed, while time-varying covariates, $\bm{X}_t$, are partially missing. This situation is common in clinical settings: treatments are typically well-documented, and key longitudinal outcomes (e.g., disease-activity indices) are routinely monitored. Cruciallly, due to informative monitoring of patients in clinical settings, the missingness of covariates is likely to depend on their underlying  values. 
For example, in the MIMIC-III dataset, patients exhibiting symptoms suggestive of abnormal heart rates were more likely to receive frequent monitoring. Consequently, the heart rate measurements may be nonignorably missing, as their absence corresponds to periods when patients’ heart rates were perceived as clinically unremarkable.

Let $R_t$ represent the missingness indicator for $\bm{X}_t$ such that $R_{t} = 1$ if all variables in $\bm{X}_t$ are fully observed and $R_{t} = 0$ if at least one element of $\bm{X}_t$ is missing. Define $\bar{\bm{R}}_t = (R_1,\ldots, R_t)$ as the history of missingness indicators up to stage $t$. We use $\bm{1}_{t}$ to denote a $1 \times t$ vector of ones 
for $t=1, \ldots, T$. 
Let $S_{t} = \{i: \bar{\bm{R}}_t = \bm{1}_t\}$ denote the sample with complete covariates until stage $t$ and $S_{t,a_t} = \{i: \bar{R}_t = \bm{1}_t, A_t=a_t\}$ denote the subsample of $S_t$ that received treatment $a_t$ at stage $t$ for $t=1,\cdots, T$. For convenience, we use $S_0$ to denote the full sample of $n$ patients. And $n_t$ is the sample size of $S_t$.
Since the conditional distribution of $Y^{pse}_t$ given $(\bm{h}_t,a_t)$, $P(Y^{pse}_t\mid \bm{h}_t,a_t)$,  may vary by  samples, we use $Q_{t,S_t}(\bm{h}_t,a_t) = E_{ S_t} (Y^{pse}_t\mid \bm{h}_t,a_t)$ to denote the Q-function based on $P(Y^{pse}_t\mid \bm{h}_t,a_t)$ in the complete-case sample $S_t$. And the value function of a stage-$t$ decision rule $d_t$ in the complete-case sample $S_t$ is $$V_{t,S_t}(d_t) = \int_{ S_t} Q_{t,S_t}\{ \bm{h}_{t}, d_t(\bm{h}_{t})\} \mathcal{R}_t(d\bm{h}_t),$$
where $\mathcal{R}_t$ is a reference distribution for $\bm{h}_t$.


In ICU settings, the effect of a treatment is commonly evaluated by the patient's longitudinal outcomes (e.g., physiological symptoms) observed several hours after treatment administration, while the treatment decision for a patient is based on the patient's medical history. Because of the temporal ordering of these variables, it is plausible that the missingness indicators $\bar{\bm{R}}_t$ do not directly depend on the patient's variables measured after receiving the current stage treatment $A_t$. Therefore, following \cite{sun2025weighted}, we make the \textit{future-independent missingness} assumption:
\begin{description}
    \item {\em Assumption} 4 (Future-independent missingness). $\bar{\bm{R}_t} \perp\!\!\!\perp (Y_t, \bm{X}_{t+1}, R_{t+1},  \ldots, \bm{X}_T,R_T,A_T,Y_T) \mid (\bm{H}_t, A_t)$  for $t=1,\ldots,T$.\label{A4}
\end{description}
Under Assumption 4, we have $Y^{pse}_t \perp\!\!\!\perp \bar{\bm{R}}_t \mid (\bm{H}_t, A_t) $ since,  given $\bm{H}_t$ and $A_t$, the pseudo-outcome $Y^{pse}_t$ only depends on $Y_t$ and the variables measured after stage $t$. Therefore, we have $P(Y^{pse}_t\mid \bm{h}_t,a_t) = P(Y^{pse}_t\mid \bm{h}_t,a_t, \bar{\bm{R}}_t = \bm{1}_t)$ and $E(Y^{pse}_t\mid \bm{h}_t,a_t) = E(Y^{pse}_t\mid \bm{h}_t, a_t, \bar{\bm{R}}_t = \bm{1}_t)$ for any given $(\bm{h}_t,a_t)$. Thus $Q_{t,S_t}(\bm{h}_{t}, a_t) = Q_{t,S_0}( \bm{h}_{t}, a_t)$. For any $\bm{h}_t \in  \mathcal{H}_t$, we have
\begin{align*}
    d^{opt}_{t,S_0}(\bm{h}_t) &= 2\mathbbm{I}\{Q_{t,S_0} (\bm{h}_{t}, 1) \geq Q_{t,S_0} (\bm{h}_{t}, -1)  \} -1 \\
    & = 2\mathbbm{I}\{Q_{t,S_t} (\bm{h}_{t}, 1) \geq Q_{t,S_t} (\bm{h}_{t}, -1)  \} -1 \\
    & = d^{opt}_{t,S_t} (\bm{h}_t).
\end{align*}

Therefore, $d^{opt}_{t,S_0}(\cdot) = d^{opt}_{t,S_t}(\cdot)$ and we can obtain the optimal stage $t$ decision rule in the full sample by estimating the optimal decision rule in the complete-case sample, that is, the sample with complete covariate data before making the treatment decision at stage $t$ (i.e., $\bar{\bm{R}}_t = \bm{1}_t$). 


\subsection{Robust estimation of optimal single-stage  rule with nonignorable missing covariates} \label{singlestage}

To convey the idea, in this section we describe the proposed direct-search methods for the optimal single-stage rule before delving into the method for the optimal multiple-stage rule in Section~\ref{multistage}.

\subsubsection{Direct-search methods and inverse probability weighting }

\

Direct-search methods typically involve a two-step procedure: (1) estimating the mean outcome for a fixed treatment rule, and (2) applying search or optimization techniques to identify the optimal treatment rule. Many classical causal inference methods have been utilized to implement Step (1). For example, in the single-stage scenario, \cite{zhao2012estimating} employed  inverse probability weighting (IPW)  to estimate the value of a  treatment rule in the full sample ($S_0$) with the following estimator
\begin{align*}
    \hat{V}_{1,S_0}(d_1) = \hat{E}_{i \in S_0} \left[ \frac{\mathbbm{I}\left\{A_{1,i}=d_1(\bm{h}_{1,i})\right\}y_{1,i}}{\mathbbm{I}(A_{1,i}=1)\hat{P}_{S_0}(A_{1}=1 \mid \bm{h}_{1,i})+\mathbbm{I}(A_{1,i}=-1)\hat{P}_{S_0}(A_{1}=-1 \mid \bm{h}_{1,i})}      \right],
\end{align*}
where $\hat{E}_{i \in S_0}[\cdot]$ denotes the sample average in  $S_0$, $\hat{P}_{S_0}(A_{1}=a_{1}\mid \bm{h}_{1})$ ($a_1 \in \{-1,1\}$) denotes the estimated propensity score of treatment  in $S_0$. The optimal treatment rule is estimated by maximizing $\hat{V}_{1,S_0}(d_1)$. Under Assumption 4, $d^{opt}_{1,S_0} = d^{opt}_{1,S_1}$ and we can obtain consistent estimates of $d^{opt}_{1,S_0}$ using the complete-case sample $S_1$.


In practice, the performance of IPW is often compromised by unstable and extreme weights estimated by prediction models for treatment \citep{kang2007demystifying}. These issues can undermine the robustness of direct-search methods involving IPW.   As an alternative, balancing weights methods find weights by directly optimizing covariate balance across treatment groups, which has been shown theoretically and empirically to improve the robustness of inverse probability weighted estimators \citep{Chattopadhyay2020}.  The main idea is to estimate weights by enforcing moment conditions that align the weighted covariate distributions of treatment groups with those of the overall study population.
To understand this, note that 
\begin{align}
    E\left\{  \frac{\mathbbm{I}(A_1=a_1)q_1(\bm{H}_1)}{P(A_1=a_1\mid \bm{H}_1)}  \right\} = E\{q_1(\bm{H}_1)\} \label{W3_eq1}
\end{align}
for  $a_1 \in \{-1,1\}$ and any moment function $q_1(\cdot)$ such that $E\{q_1(\bm{H}_1)\}$ exists and is finite. Because the purpose of IPW is to achieve covariate balance across treatment groups after weighting, instead of predicting the treatment propensity score, it is natural to find weights $w_1$ that satisfy the empirical finite-dimensional approximation of ~\eqref{W3_eq1},
\begin{align}
    \hat{E}_{i \in S_1} \left\{  w_{1,i}\mathbbm{I}(A_{1,i}=a_{1})\tilde{\bm{q}}_1(\bm{h}_{1,i}) \right\} = \hat{E}_{i \in S_1} \{\tilde{\bm{q}}_1(\bm{h}_{1,i})\}, a_1 \in \{-1,1\}, \label{W3_eq2}
\end{align}
where $\tilde{\bm{q}}_1(\cdot)$ is composed of finite-order moments functions. While balancing lower-order moments of the covariates often improves the performance of weighting,  this does not guarantee adequate balance across a broader class of functions spanning the potential outcome mean space. Building upon \cite{wong2018kernel}, we propose to enforce covariate functional balance over a reproducing-kernel Hilbert space (RKHS) of functions in order to improve weight estimation in Step (1) of the direct-search methods. 

\subsubsection{Balancing weights for  potential outcome means}

\

Let $Q_{1,a_1}(\bm{h}_1) = Q_{1}(\bm{h}_1,a_1) =  E\{Y^*_1(a_1)\mid \bm{h}_1\}$. Suppose that $Q_{1,1}(\cdot)$ belongs to a RKHS $\mathcal{Q}_{1,1}$ with inner product $\langle \cdot,\cdot \rangle_{\mathcal{Q}_{1,1}}$ and norm $\| \cdot \|_{\mathcal{Q}_{1,1}}$. 
Similarly, we assume $Q_{1,-1}$ belongs to a RKHS $\mathcal{Q}_{1,-1}$. 
For any function $q_{1,1}(\cdot) \in \mathcal{Q}_{1,1}$, to balance its distribution between the  complete-case  sample (i.e., $S_1$) and its subset with $A_1=1$ (i.e., $S_{1,1}$), it is natural to choose weights that minimize the imbalance  measure
\begin{align}
    F^{imb}(\bm{w}_1,q_{1,1}) = \left( \hat{E}_{i \in S_1}\left[ \{\mathbbm{I}(A_{1,i}=1) w_{1,i}-1\}q_{1,1}(\bm{h}_{1,i})  \right] \right)^2 , \label{imbfun}
\end{align}
where $w_{1,i}$ is the balancing weight for patient $i$ in the complete-case  sample $S_1$ and $\bm{w}_1$ denotes the vector of the balancing weights. $F^{imb}(\bm{w}_1,q_{1,1})$ summarizes the imbalance of the values of  $q_{1,1}(\bm{h}_{1})$ between  $S_{1,1}$ and $S_{1}$  after weighting by $\bm{w}_1$.


Let $p_1$ be  the dimension of $\bm{H}_1$. Suppose the support of the  distributions of $\bm{H}_1$ is $[0,1]^{p_1}$, and $q^{(\zeta)}$ is the $\zeta$th derivative of a function $q$. Following \cite{wong2018kernel}, we choose $\mathcal{Q}_{1,1}$ as the tensor product RKHS $\mathcal{W}_{1}^{\zeta_{1,1},2} \bigotimes \mathcal{W}_{2}^{\zeta_{1,1},2}\bigotimes \ldots \mathcal{W}_{p_1}^{\zeta_{1,1},2}$ with $\mathcal{W}_j^{\zeta_{1,1},2}$ being the $\zeta_{1,1}$th-order Sobolev space of functions of the $j$th component of $\bm{h}_1$ and $\mathcal{W}^{\zeta_{1,1},2}([0,1]) = \{q: q, q^{(1)},\ldots, q^{(\zeta_{1,1}-1)} $ are absolutely continuous, $ q^{(\zeta_{1,1})}\in L^2[0,1] \}$ with norm 
$\|q\|_{\mathcal{Q}_{1,1}} = \left[ \sum_{k=0}^{\zeta_{1,1}-1}\left\{ \int_0^1 q^{(k)}(x)dx \right\}^2  + \int_0^1 \left\{ q^{(\zeta_{1,1})}(x) dx \right\}^2   \right]^{\frac{1}{2}}$.


Ideally, we can optimize the balance of the covariate functionals between the  samples $S_{1,1}$ and $S_1$ by controlling $\sup_{q_{1,1}\in \mathcal{Q}_{1,1}} F^{imb}(\bm{w}_1,q_{1,1}) $. 
However,  $F^{imb}(\bm{w}_1,q_{1,1}) $ faces a scale issue with respect to $q_{1,1}$, and it is necessary to alleviate the unstableness of $\bm{w}_1$ using penalty terms \citep{wong2018kernel}. Therefore, following \cite{wong2018kernel}, we set the target function to minimize as 
\begin{align}
    L^{bal}_{1,1}(\bm{w}_1) = \sup\limits_{q_{1,1} \in \mathcal{Q}_{1,1}} \left\{  \frac{F^{imb}(\bm{w}_1,q_{1,1})}{\| q_{1,1} \|_{n_1}^2} - \lambda^{bal,1}_{1,1} \frac{\| q_{1,1} \|_{\mathcal{Q}_{1,1}}^2}{\| q_{1,1} \|_{n_1}^2} \right\} + \lambda^{bal,2}_{1,1} J_{1,1}^{bal}(\bm{w}_1) \label{bal_loss_fun},
\end{align}
where  $\|q_{1,1}\|_{n_1}^2 = \hat{E}_{i \in S_1}\{q_{1,1}(\bm{h}_{1,i})^2\} = 1$ to avoid the scale issue, 
 $\| \cdot \|_{\mathcal{Q}_{1,1}}$ is controlled with the tuning parameter $\lambda^{bal,1}_{1,1}>0$ to emphasize the balance on smoother functions, and  $J_{1,1}^{bal}(\bm{w}_1) = \hat{E}_{i\in S_1}\left\{\mathbbm{I}(A_{1,i}=1) w_{1,i}^2\right\}$ is penalized with the tuning parameter $\lambda^{bal,2}_{1,1}>0$ to control the variability of $\bm{w}_1$.
 The weights are also restricted to be greater than or equal to 1, as their counterparts, the inverse propensity scores of treatment, are no less than 1. By the representer theorem of \cite{wahba1990spline}, the solution to the minimization of~\eqref{bal_loss_fun} enjoys a finite-dimensional representation, even if 
  $\mathcal{Q}_{1,1}$ is infinite-dimensional.
 Details of the construction and minimization of~\eqref{bal_loss_fun}, including tuning parameter selection,  are provided in Appendix B of the Supplementary Materials.

Note that the minimization of $L^{bal}_{1,1}(\bm{w}_1)$ is only taken over $w_{1,i}(i: i\in S_{1,1})$. 
In a similar manner, we can construct $L^{bal}_{1,-1}(\bm{w}_1)$ and its minimization is taken over $w_{1,i}(i: i\in S_{1,-1})$. By combining these estimated weights across mutually exclusive patient treatment groups, we obtain the estimated balancing weights for the entire $S_1$ sample, which is denoted as $\hat{\bm{w}}^{b}_{1}.$

\subsubsection{Balancing weights estimators of treatment regime values}

\

Let $\hat{\Omega}^{BW}_{1,i}(a_1) = \hat{w}^{b}_{1,i}\mathbbm{I} (A_{1,i}=a_1)y_{1,i} $ for $a_1\in \{-1,1\}$. We construct the estimator of $V_{1,S_1}(d_1)$ with the balancing weights as 
\begin{align}
    \hat{V}_{1,S_1}^{BW}(d_1) =\hat{E}_{i \in S_1} \left[ \hat{\Omega}^{BW}_{1,i}\{d_1(\bm{h}_{1,i})\}  \right].
\end{align}

For efficient estimation of $V_{1,S_1}(d_1)$, we also propose an augmented balancing weights estimator using additional regression models for $Q_{1,1}(\cdot)$ and $Q_{1,-1}(\cdot)$. 
Since  $Q_{1,1}(\cdot)$ and  $Q_{1,-1}(\cdot)$ are assumed to be in RKHS, it is natural to employ reproducing-kernel-based methods (e.g., smoothing splines) to estimate them when the following assumption holds in the single-stage scenario (i.e., $T=1$).
\begin{description}
    \item {\em Assumption} 5. For $t=T, \ldots, 1$, $Y^{pse}_t = Q_{t}(\bm{h}_t,a_t)+\epsilon_t$. The errors $\{ \epsilon_{t,i} \}$ are uncorreleated, with $E(\epsilon_{t,i})=0$ and 
    $var(\epsilon_{t,i}) = \sigma_{t,i}^2\leq \sigma_t^2$ for some constant $\sigma_t^2$,  $\forall i\in S_t$. Further, $\{\epsilon_{t,i}\}$ are independent of $A_{t,i}$ and $\bm{H}_{t,i}$.
\end{description} 

We provide the details for estimating $Q_{1,1}(\cdot)$ and $Q_{1,-1}(\cdot)$ using smoothing splines in Appendix C of the Supplementary Materials.
Let $\hat{Q}_{1,1}(\cdot)$ and $\hat{Q}_{1,-1}(\cdot)$ denote smoothing spline estimators  for $Q_{1,1}(\cdot)$ and and $Q_{1,-1}(\cdot)$, respectively. Then the Q-function estimator  is $\hat{Q}_1(\bm{h}_1,a_1) = \hat{Q}_{1,a_1}(\bm{h}_1)$. Let $\hat{\Omega}^{ABW}_{1,i}(a_1) = \hat{w}^b_{1,i}\mathbbm{I}(A_{1,i}=a_1)y_{1,i} - \{\hat{w}^b_{1,i}\mathbbm{I}(A_{1,i}=a_1)-1\} \hat{Q}_{1}(\bm{h}_{1,i},a_1)$ for $a_1\in \{-1,1\}$. The augmented balancing weighting estimator of $V_{1,S_1}(d_1)$ is 
\begin{align*}
    \hat{V}_{1,S_1}^{ABW}(d_1) =\hat{E}_{i \in S_1} \left[ \hat{\Omega}^{ABW}_{1,i}\{d_1(\bm{h}_{1,i})\}  \right].
\end{align*}

\subsubsection{Estimation of optimal treatment rules}

\

Let $\mathcal{G}_1$ be the class of measureable functions from $\mathbbm{R}^{p_1}$ into $\mathbbm{R}$. Any decision rule $d_1(\bm{h}_1)$ can be written as $d_1(\bm{h}_1) = \text{sgn}\{ g_1(\bm{h}_1) \}$ for some function $g_1 \in \mathcal{G}_1$, where we define $\text{sgn}(0)=1$. Thus, we  focus on the estimation of $g_1$ within a class of functions $\mathcal{G}_1$ called the approximation space. As shown by \cite{zhao2019efficient}, the decision rule that optimize $\hat{V}_{1,S_1}(d_1)$ (including $\hat{V}^{BW}_{1,S_1}(d_1)$ and $\hat{V}^{ABW}_{1,S_1}(d_1)$) can be obtained by minimizing
\begin{align*}
  \hat{E}_{i \in S_1}\left( |\hat{\Omega}_{1,i}(1)|\mathbbm{I}\left[ \text{sgn}\{\hat{\Omega}_{1,i}(1)\}g_1(\bm{h}_{1,i})<0 \right] + |\hat{\Omega}_{1,i}(-1)|\mathbbm{I}\left[ -\text{sgn}\{\hat{\Omega}_{1,i}(-1)\}g_1(\bm{h}_{1,i})<0 \right] \right),
\end{align*}
where $\hat{\Omega}_{1,i}$ can be $\hat{\Omega}^{BW}_{1,i}$ or $\hat{\Omega}^{ABW}_{1,i}$.
This can be viewed as minimizing a sum of weighted 0-1 losses, which is a difficult non-convex optimization problem. Therefore, following \cite{zhao2019efficient}, we replace the indicator function with a convex surrogate function and minimize the resulting relaxed objective function. Let $\phi(\cdot): \mathcal{R}\to\mathcal{R}$ denote a convex function and define the 
estimators of optimal treatment rules as those taking the form 
\begin{align*}
      \arginf\limits_{g_1\in \mathcal{G}_1} \hat{E}_{i \in S_1}\left( |\hat{\Omega}_{1,i}(1)|\phi\left[ \text{sgn}\{\hat{\Omega}_{1,i}(1)\}g_1(\bm{h}_{1,i})\right] + |\hat{\Omega}_{1,i}(-1)|\phi\left[ -\text{sgn}\{\hat{\Omega}_{1,i}(-1)\}g_1(\bm{h}_{1,i})\right ] \right) +\lambda^{opt}_1 \|g_1\|^2,
\end{align*}
where 
$\lambda^{opt}_1 \|g_1\|^2$ is included to reduce over-fitting and $\lambda^{opt}_1$ is a (possibly data-dependent) tuning parameter. Following \cite{zhao2019efficient}, we take $\phi(x) = \log (1+e^{-x})$ to employ the logistic loss as the surrogate for 0-1 loss.  When we let $\hat{\Omega}_{1,i}=\hat{\Omega}^{BW}_{1,i}$, the optimal rule estimator is called `covariate-functional-balancing learning' (CFBL) estimator. And when we let $\hat{\Omega}_{1,i}=\hat{\Omega}^{ABW}_{1,i}$, it is called the `augmented covariate-functional-balancing learning' (ACFBL) estimator.

The estimation procedure is summarized as follows:

\begin{enumerate}
    \item Obtain the balancing weights $\hat{\bm{w}}^{b}_{1}$ for the complete-case sample $S_1$ by minimizing $L_{1,1}^{bal}(\bm{w}_1)$ and $L_{1,-1}^{bal}(\bm{w}_1)$. 
    \item Estimate $Q_{1,1}(\bm{h}_1)$ and $Q_{1,-1}(\bm{h}_1)$ with reproducing-kernel-based estimators (e.g., smoothing splines), denote the estimators as $\hat{Q}_{1,1}(\bm{h}_1)$ and $\hat{Q}_{1,-1}(\bm{h}_1)$.
    \item Construct $\hat{\Omega}_{1,i}$ using
    $\hat{\Omega}^{BW}_{1,i}(a_1)$ or $\hat{\Omega}^{ABW}_{1,i}(a_1)$
    for $a_1\in \{1,-1\}$. Note that Step (2) is skipped if using $\hat{\Omega}^{BW}_{1,i}(a_1)$.
    \item Obtain $\hat{g}_1$ by minimizing $$\hat{E}_{i \in S_1}\left( |\hat{\Omega}_{1,i}(1)|\phi\left[ \text{sgn}\{\hat{\Omega}_{1,i}(1)\}g_1(\bm{h}_{1,i})\right] + |\hat{\Omega}_{1,i}(-1)|\phi\left[ -\text{sgn}\{\hat{\Omega}_{1,i}(-1)\}g_1(\bm{h}_{1,i})\right ] \right) +\lambda^{opt}_1 \|g_1\|^2.$$ 
    \item $\hat{d}^{opt}_1(\bm{h}_1) = \text{sgn}\{\hat{g}_1 (\bm{h}_1) \}$.
\end{enumerate}


Proofs of the consistency of $\hat{V}^{BW}_{1,S_1}(d_1)$, $\hat{V}^{ABW}_{1,S_1}(d_1)$, and $\hat{d}^{opt}_1(\bm{h}_1)$ are provided in Appendix D of the Supplementary Materials.

\subsection{Robust estimation of optimal multi-stage  rule with nonignorable missing covariates}\label{multistage}

\subsubsection{Motivation of backward induction using pseudo-outcomes}
In multiple-stage scenarios, the optimal treatment decision at any given stage must take into account the decision rules and potential outcomes of future stages.  To address this, two approaches are commonly used within the direct-search framework, depending on whether Q-functions are employed.  
The first approach estimates the value function at stage $t$ by weighting observed (current and future) rewards using the inverse of treatment propensity scores for the future stages \citep{zhao2015new,liu2018augmented}. The second approach estimates the value function under a multi-stage rule by employing Q-functions and estimated potential outcome means in addition to treatment propensity scores \citep{Schulte_et_al_2014,zhang2013robust,zhang2018c}.


With nonignorable missing covariates, these two approaches present different challenges for estimating optimal multi-stage rules.
Under the future-independent missingness assumption, we have $d_{t,S_0}^{opt}(\cdot) = d_{t,S_t}^{opt}(\cdot)$ for stage $t = T-1,\ldots,1$.  However, in the first approach, the value function estimation at stage $t$ relies on   $P(A_{t+1}\mid \bm{h}_{t+1})\ldots P(A_T\mid \bm{h}_T)$, and thus would be influenced by covariate missingness in all propensity score models of stages $t+1, \ldots, T$. As a result, the first approach only utilizes observed data from patients with complete covariate data \textit{throughout all stages}, which exacerbates the efficiency concerns of IPW-based methods.   In contrast, the second approach also employs $Q_{t+1}(\bm{h}_{t+1},a_{t+1})$ to estimate the potential outcome means, which is only affected by the missingness of $\bm{H}_{t+1}$. Therefore, information from patients with missing covariate values at stages $t+2, \ldots, T$ is used for the value function estimation at stage $t$, which could provide efficiency gains over the first approach in practice. Nevertheless, utilizing Q-functions also introduces the risk of model misspecification, which could be mitigated by flexible models. 
In this article, we follow the second approach and utilize  $\hat{Q}_{t+1}\{\bm{h}_{t+1},\hat{d}_{t+1}^{opt}(\bm{h}_{t+1})\}$ 
 estimated by smoothing splines as $\hat{Y}^{pse}_t$  for estimating the value function   $V_{t,S_t}(d_t)$.

\vspace{-8pt}
\subsubsection{Handling nonignorable missingness in pseudo-outcomes}

The optimal decision rule at stage $t$ can be estimated using  $(\bm{H}_t,A_t,Y^{pse}_t)$ in $S_t$, similar to the optimal single-stage decision rule estimation  in Section~\ref{singlestage} that employs  $(\bm{H}_1,A_1,Y_1)$ in $S_1$, provided no missingness exist in relevant variables. For example, at the final stage $T$, $(\bm{H}_T,A_T,Y^{pse}_T)$ are fully observed in $S_T$, allowing direct application of  the proposed methods  in Section~\ref{singlestage}. However, for stages $t=1,\ldots,T-1$,  nonignorable missingness of  $\bm{H}_{t+1}$ would propagate to the pseudo-outcome  $Y^{pse}_t$ within  $S_t$. Thus, while the estimation of the balancing weights $\hat{\bm{w}}_t^b$ remains unaffected since ${Y}^{pse}_t$ is not involved, the missingness of ${Y}^{pse}_t$ in $S_t$ creates challenges in estimating  the Q-function. 

Let $R^{pse}_t$ denote the missingness indicator of $Y^{pse}_t$ and let $Q_{t+1,a_{t+1}}(\bm{h}_{t+1})= Q_{t+1}(\bm{h}_{t+1},a_{t+1})$. $R^{pse}_t$ is determined by the missingness of  $\bm{H}_{t+1}$ and  $d^{opt}_{t+1}(\bm{h}_{t+1})$, as well as  the form of  $Q_{t+1,a_{t+1}}(\bm{h}_{t+1})$. When all members of $\bm{H}_{t+1}$ are required to compute $Y^{pse}_t$ (e.g., in linear models when all coefficients in  $Q_{t+1,a_{t+1}}(\bm{h}_{t+1})$  are non-zero),  $R^{pse}_t$ is equivalent to  $\mathbbm{I}(\bar{\bm{R}}_{t+1}=\bm{1}_{t+1})$.
Because $Y^{pse}_t$ is computed based on $\bm{H}_{t+1}$ (including $\bm{X}_{t+1}$), 
the missingness of $\bm{X}_{t+1}$ would lead to  the missingness of $Y^{pse}_t$. As a result,  $Y^{pse}_t$ could be missing even conditional on $\bar{\bm{R}}_t=\bm{1}_t$ in  $S_t$. 
Moreover, because $\bm{X}_{t+1}$ is directly associated with $R_{t+1}$, $Y^{pse}_t$ is likely to be associated with  $R_{t+1}$ even if conditional on $(\bm{H}_t,A_t)$ in $S_t$. These associations imply that  $R^{pse}_t$ is likely to depend on $Y^{pse}_t$ conditional on $(\bm{H}_t,A_t)$ in $S_t$, leading to nonignorable missingness of the pseudo-outcomes.

This nonignorable missingness issue in the pseudo-outcomes could result in biased estimates of the Q-functions if a complete-case analysis or multiple imputation is applied because $E_{S_t}(Y^{pse}_t\mid \bm{h}_t,a_t) \neq E_{S_t}(Y^{pse}_t\mid \bm{h}_t, a_t, R^{pse}_t=1)$.
To address nonignorable missing pseudo-outcomes, \cite{sun2025weighted} applied IPW to  Q-learning, where a semiparametric working model for the missingness propensity was identified and estimated with the aid of nonresponse instrumental variables. We extend the missingness propensity model of \cite{sun2025weighted} as follows: 
\begin{description}
    \item {\em Assumption} 6. (a semiparametric model for missingness propensity of the pseudo-outcomes). For $t = 1,\ldots,T-1$, 
\begin{align}\label{missingnessmodel}
    P(R^{pse}_t=1\mid \bm{h}_t,a_t,y^{pse}_t,\bar{\bm{R}}_t=\bm{1}_t) = \pi(\bm{u}_t,y^{pse}_t;\bm{\gamma}_t)= \frac{1}{1+\exp\{\eta_t(\bm{u}_t)+\Gamma_{t,\bm{\gamma}_t}( y^{pse}_t)\}},
\end{align}
\end{description}
where $\bm{U}_t \subset (\bm{H}_t,A_t)$ is a proper subset of $(\bm{H}_t,A_t)$ such that $P(Y^{pse}_t\mid \bm{h}_t, a_t) \neq P(Y^{pse}_t\mid \bm{u}_t)$, $\eta_t(\cdot)$ is an unknown and unspecified function of $\bm{u}_t$, $\Gamma_{t,\bm{\gamma}_t}(\cdot)$ is a known parametric function of $y^{pse}_t$ with an \textit{unknown} parameter vector $\bm{\gamma}_t$. $\Gamma_{t,\bm{\gamma}_t}(\cdot)$ is more general than the simple function $\gamma_ty^{pse}_t$ in  the missingness propensity model of \cite{sun2025weighted} by allowing transformations of $y^{pse}_t$ (e.g., log transformation or power terms).

Assumption 6 implies the existence of nonresponse instrumental variables at stage $t$. Specifically, under Assumption 6,  the pair $(\bm{H}_t,A_t)$ can be decomposed as $(\bm{H}_t,A_t) = (\bm{U}_t,\bm{Z}_t)$ such that $\bm{Z}_t \not\perp\!\!\!\perp Y^{pse}_t \mid \bm{U}_t$ and $\bm{Z}_t \perp\!\!\!\perp R^{pse}_t \mid  \bm{U}_t, Y^{pse}_t$ when $\bar{\bm{R}}_t=\bm{1}_t$. That is, conditional on $\bm{U}_t$, $\bm{Z}_t$ is a predictor vector of the pseudo-outcome $Y^{pse}_t$, but given $\bm{U}_t$ and $Y^{pse}_t$, $\bm{Z}_t$ is conditionally independent of the missingness indicator $R^{pse}_t$.
$\bm{Z}_t$ is referred to as the nonresponse instrumental variables at stage $t$, which allow the identification and estimation of $\bm{\gamma}_t$ in the working model~\eqref{missingnessmodel} using estimating equations \citep{Shao_Wang_2016,Miao_Ding_Geng_2016,li2023non}. 
In Section~\ref{application}, we will provide a detailed discussion on the plausibility of Assumption 6 in the context of the MIMIC-III data analysis.

Following  \cite{Shao_Wang_2016} and \cite{sun2025weighted}, we apply  the idea of profiling to obtain the following kernel-regression estimate of $\eta_t(\cdot)$ for  fixed values of  $\bm{\gamma}_t$,
\begin{align}
    \exp\{\hat{\eta}_{\bm{\gamma}_t,t}(\bm{u}_t) \} =  \frac{\sum_{i=1}^{n} \left(1-r^{pse}_{t,i}\right)  K^m_{c_t}\left(\bm{u}_t-\bm{u}_{t,i}\right) \mathbbm{I}(\bar{\bm{R}}_{t,i}=\bm{1}_t) }{\sum_{i=1}^{n}  r^{pse}_{t,i}   \exp \left\{\Gamma_{t,\bm{\gamma}_t}(y^{pse}_{t,i})\right\} K^m_{c_t}\left(\bm{u}_t-\bm{u}_{t,i}\right) \mathbbm{I} (\bar{\bm{R}}_{t,i}=\bm{1}_t)}, \label{missingkernel}
\end{align}
where $K^m_{c_t}(\cdot) = c_t^{-1}K^m(\cdot/c_t)$, with $K^m(\cdot)$ being a symmetric kernel function and $c_t$ a bandwidth. It is notable that although kernels are used in the estimation of both the balancing weights and missingness propensity models, they act in different roles. When estimating the balancing weights, the reproducing kernels are used for obtaining a finite representation for the infinite functional RKHS, whereas for estimating the nonparametric form of $\exp\{\hat{\eta}_{\bm{\gamma}_t,t}(\bm{u}_t)\}$, the kernel regression method is used as a local fitting technique.

Let $\hat{\pi}_t(\bm{u}_t,y^{pse}_{t};\bm{\gamma}_t) = [1+\exp\{\hat{\eta}_{\bm{\gamma}_t,t}(\bm{u}_t) + \Gamma_{t,\bm{\gamma}_t}(y^{pse}_t) \} ]^{-1}$. Then, we can estimate $\bm{\gamma}_t$ by solving the following estimating equation: 
\begin{align}
    \widehat{E} \left[ \mathbbm{I} (\bar{\bm{R}}_t=\bm{1}_t) \bm{l}_t(\bm{z}_t)\left\{\frac{r^{pse}_t}{\hat{\pi}_t(\bm{u}_t,y^{pse}_{t}; \bm{\gamma}_t)} -1 \right\} \right]  = \bm{0}, \label{missingEE}
\end{align}  where $\bm{l}_t(\bm{z}_t)$ is a user-specified differentiable vector function of the nonresponse instrumental variables. 
Let $dim(\cdot)$ denote the dimension of a vector. If $dim\left[\bm{l}_t(\bm{z}_t)\left\{\frac{r^{pse}_t}{\hat{\pi}_t(\bm{u}_t,y^{pse}_{t}; \bm{\gamma}_t)}\right\}\right] < dim(\bm{\gamma}_t)
+1$, the equations in ~\eqref{missingEE} are not solvable because they are under-identified. Nevertheless, following \cite{shao2016semiparametric}, when $dim\left[\bm{l}_t(\bm{z}_t)\left\{\frac{r^{pse}_t}{\hat{\pi}_t(\bm{u}_t,y^{pse}_{t}; \bm{\gamma}_t)}\right\}\right] > dim(\bm{\gamma}_t)+1$, 
 we can employ the two-step generalized method of moments (GMM) \citep{hansen1982large} to estimate $\bm{\gamma}_t$. Details are provided in Appendix E of the Supplementary Materials. 
With the estimated missingness propensity $\hat{\pi}_t$ from  the model in~\eqref{missingnessmodel}, we can estimate ${Q}_t(\bm{h}_t,a_t)$ by weighted smoothing splines. Details are provided in Appendix F of the Supplementary Materials.

With $\hat{\bm{w}}^b_t$, $\hat{\pi}_t$  and $\hat{Q}_t(\bm{h}_t,a_t)$, the augmented balancing weights estimator for $V_{t,S_t}(d_t)$ can be constructed as 
\begin{align*}
    \hat{V}_{t,S_t}^{ABW}(d_t) = \hat{E}_{i\in S_t}\left\{ \frac{r^{pse}_{t,i}}{\hat{\pi}_{t,i}} \left[ \hat{w}^b_{t,i}\mathbbm{ I}\{A_{t,i}=d_t(\bm{h}_{t,i})\}\hat{y}^{pse}_{t,i} - [\hat{w}^b_{t,i}\mathbbm{I}\{A_{t,i}=d_t(\bm{h}_{t,i})\}-1] \hat{Q}_{t}\{\bm{h}_{t,i},d_t(\bm{h}_{t,i})\}    \right] \right\}.
\end{align*}
The minimization of $\hat{V}_{t, S_t}^{ABW}(d_t)$ to obtain the optimal decision rule function $\hat{d}^{ABW}_t(\cdot)$ is achieved by the same approach as in the single-stage scenario. 

\subsubsection{Summary of the estimation procedure}
We summarize the estimation procedure of the optimal multiple-stage rule. For $t = T,\ldots, 1$, repeat the following steps recursively:
\begin{enumerate}
    \item Minimize $L^{bal}_{t,1}$ and $L^{bal}_{t,-1}$ and obtain $\hat{\bm{w}}^b_t$ for the $S_t$ sample.
    \item Construct $\hat{Y}^{pse}_{t}$:
    \begin{enumerate}[(1).]
        \item When $t=T$, $\hat{Y}^{pse}_{t} = f(Y_1,\ldots,Y_T)$.
        \item When $t \in \{T-1,\ldots,1\}$, $\hat{Y}^{pse}_{t} =\hat{Q}_{t+1}\{\bm{H}_{t+1},\hat{d}_{t+1}^{opt}(\bm{H}_{t+1})\}$.
    \end{enumerate}
    \item Estimate $\hat{\pi}_t$:
    \begin{enumerate}[(1).]
        \item When $t=T$, $\hat{\pi}_t$ = 1.
        \item When $t \in \{T-1,\ldots,1\}$, 
        obtain $\hat{\eta}_{\bm{\gamma}_t,t}(\bm{u}_t)$ by~\eqref{missingkernel} and then estimate $\bm{\gamma}_t$ by solving the estimating equation~\eqref{missingEE}. Then, $\hat{\pi}_t$ can be estimated by plugging in $\hat{\eta}_{\hat{\bm{\gamma}}_t,t}(\bm{u}_t)$ and $\hat{\bm{\gamma}}_t$ in model~\eqref{missingnessmodel}. 
    \end{enumerate}
    \item 
    Obtain   $\hat{Q}_{t}(\bm{h}_{t},a_t)$ ($a_t\in \{-1,1\}$) by weighted smoothing splines with inverse weights $1/\hat{\pi}_t$.  
    \item Construct $\hat{\Omega}_{t,i}(a_t) $ as $  \hat{w}^b_{t,i}\mathbbm{I}\{A_{t,i}=a_t\}\hat{y}^{pse}_{t,i} - [\hat{w}^b_{t,i}\mathbbm{I}\{A_{t,i}=a_t\}-1] \hat{Q}_{t}(\bm{h}_{t,i},a_t)$ for $a_t\in \{-1,1\}$. 
    \item Obtain $\hat{g}_t(\cdot)$ by minimizing $$\hat{E}_{i \in S_t}\left[ \frac{r^{pse}_{t,i}}{\hat{\pi}_{t,i}}|\hat{\Omega}_{t,i}(1)|\phi[ \text{sgn}\{\hat{\Omega}_{t,i}(1)\}g_t(\bm{h}_{t,i})] + \frac{r^{pse}_{t,i}}{\hat{\pi}_{t,i}}|\hat{\Omega}_{t,i}(-1)|\phi[ -\text{sgn}\{\hat{\Omega}_{t,i}(-1)\}g_t(\bm{h}_{t,i}) ] \right] +\lambda^{opt}_t \|g_t\|^2.$$
    \item Obtain the optimal decision rule at stage $t$ by $\hat{d}^{opt}(\cdot) = \text{sgn}\{\hat{g}_t(\cdot)\}$.
\end{enumerate}
Since estimating equations (EE) were employed to estimate the missingness propensity of the pseudo-outcomes, we call this optimal multi-stage rule estimator the  `EE-ACFBL' estimator to distinguish it from the ACFBL estimator of the single-stage rule.  
The consistency of the EE-ACFBL estimator can be established under suitable regularity conditions for estimating equations, the functional space for potential outcome mean functions, and an additional assumption that the optimal treatment is unique for all patients at all stages. Note that the parameters are estimated for each stage separately in a recursive manner. Thus, it is natural to establish the convergence properties of the stage-specific decision rules recursively as well. For simplicity, we will focus on the two-stage setting, but extensions to the general $T$-stage setting follow directly.


Proofs of the consistency of the estimated pseudo-outcomes and optimal two-stage treatment rule are provided in Appendix G of the Supplementary Materials. We provide an \texttt{R Markdown} tutorial at \url{https://github.com/AlbertSun0930/Code-for-the-robust-DTR-MNAR-paper} to demonstrate the implementation of the EE-ACFBL estimator using simulated data.

\section{Simulation}\label{simulation}

We conducted three simulation studies to evaluate the finite-sample performance of the proposed methods, in comparison to several existing methods in scenarios with nonignorable missing covariates. In Simulation 1, we assessed the performance of the CFBL and ACFBL estimators in the single-stage scenario. In Simulation 2, we evaluated the EE-ACFBL estimator in the multiple-stage scenario. In Simulation 3, we investigated the robustness of the EE-ACFBL estimator  under mild model misspecification of the missingness propensity model (i.e., violations of Assumption 6). 
The performance of the methods was evaluated based on  (1) the values of the estimated treatment regimes and (2) the proportion of patients who have received the true optimal treatment under the estimated regime.

\subsection{Simulation 1. Single-stage scenario}

We simulated data from an observational study where Assumptions 1-6 were satisfied in the single-stage scenario. There were two covariates $X_{1,1}$ and $X_{1,2}$. $X_{1,2}$ was partially missing and its missingness indicator is $R_1$. Denote $\text{expit}(b) = \exp(b)/\{1+\exp(b)\}$ for $b$ in $\mathcal{R}$. The detailed set-up is provided in Table~\ref{simu1_DGM}.
\begin{table}[htbp]
    \centering
    \caption{Details of the  Simulation 1 set-up}
    \begin{tabular}{rcl}
    \hline
    \multicolumn{3}{l}{\textbf{Data-generating mechanism}}\\
    $X_{1,1}$  &  $\sim$ & $N(0,1) $\\
    $X_{1,2}$ & $\sim$ & $\text{Uniform}(0,2)$\\
    $R_1$  & $\sim$ & $\text{Bernoulli} ([1+\exp(-3+X_{1,2})]^{-1})$\\
    $A_1$ &  $\sim$ & $2\times \text{Bernoulli}(\text{expit}(-1+2X_{1,1}^2-X_{1,2}^2-R_1)) - 1$\\
    $Y_1$ & $=$  & $-2+2A_1(1-X_{1,2})+2 X_{1,1}^2 + X_{1,2} + \epsilon_1 ; \epsilon_1 \sim N(0,1)$\\
     \multicolumn{3}{c}{}\\
     \multicolumn{3}{l}{\textbf{Final outcome}}\\
    $Y$  & =& $Y_1$ \\
    \multicolumn{3}{c}{}\\
    \multicolumn{3}{l}{\textbf{Settings implied by the data-generating mechanism}}\\
    {$d_1^{opt}(\bm{h}_1)$} & = & $2\mathbbm{I}(1-x_{1,2})-1$ \\
    \hline
    \end{tabular}
    \label{simu1_DGM}
\end{table}

Note that  $R_1$  was  directly associated with the value of $X_{1,2}$, thus $X_{1,2}$ was nonignorably missing, and the future-independent missingness assumption held.
 We generated 1000 data sets with sample sizes of $n=500$ and $n=2000$. The missingness proportion was approximately 13.2\%.

We evaluated the following methods for estimating the optimal single-stage rule. (1) CC-QL (I): Q-learning based on incorrectly specified parametric Q-function using the complete-case (CC) sample (i.e., $S_1$). (2)  CC-OWL (N): outcome weighted learning  by \cite{zhao2012estimating} with nonparametric treatment propensity score estimation applied to $S_1$. (3) CC-EARL (II): the EARL method  by \cite{zhao2019efficient} based on incorrectly specified parametric Q-function and treatment propensity score models applied to $S_1$. (4) CC-EARL (NN): the EARL method  based on nonparametric Q-function and treatment propensity score estimation applied to $S_1$. (5) CC-CFBL: the proposed covariate functional balancing learning using $S_1$. (6) CC-ACFBL: the proposed augmented covariate functional balancing learning using $S_1$. 
(7) All-QL (C):  Q-learning based on correctly specified Q-function with \textit{complete data} (benchmark).
In the above methods, when Q-function was misspecified, it was misspecified as $Q(\bm{h}_1,a_1) = \beta_{10} + a_1(\psi_{10}+\psi_{11}x_{1,2}) + \beta_{1,2}x_{1,1}$. When the treatment propensity score model was misspecified, it was misspecified as $P(A_1=1\mid \bm{h}_1) = \text{expit}(\alpha_{10} + \alpha_{11}x_{1,1} + \alpha_{22}x_{1,2})$. Additionally, when we applied nonparametric estimation to existing methods, the Q-function was estimated using the random forest method via the R package \texttt{randomForest} with the default hyperparameter set-up. Similarly, the nonparametric estimation of the treatment propensity score was implemented using the support vector machine  method via the R package \texttt{e1071}, also with the default hyperparameter set-up. For all methods, the decision set was always correctly specified as $\mathbbm{I}(\psi_{10}+\psi_{11}x_{1,2}>0)$.

\begin{table}[htbp]
  \centering
  \caption{Performance of the estimated optimal single-stage rules in Simulation 1}
  \begin{threeparttable}
    \begin{tabular}{lcccc}
    \hline
          & \multicolumn{2}{c}{n=500}     & \multicolumn{2}{c}{n=2000} \\
          \cmidrule(lr){2-3}\cmidrule(lr){4-5}
          & Value & Opt\% & Value & Opt\% \\
    \hline
    CC-QL (I) & 1.420 (0.175) & 0.624 (0.065) & 1.426 (0.095) & 0.623 (0.037)\\
    CC-OWL (N) & 1.930 (0.129) & 0.879 (0.053) & 1.953 (0.084) & 0.900 (0.040) \\
    CC-CFBL & 1.984 (0.126) & 0.947 (0.032) & 1.994 (0.065) & 0.965 (0.017)\\
    CC-EARL (II) & 1.093 (0.169) & 0.525 (0.047) & 1.035 (0.086) & 0.509 (0.023) \\
    CC-EARL (NN) & 1.949 (0.127) & 0.899 (0.048) & 1.976 (0.072) & 0.929 (0.030) \\
    CC-ACFBL & 1.989 (0.128) & 0.958 (0.027) & 1.995 (0.065) & 0.971 (0.017) \\
    All-QL (C) & 2.009 (0.127) & 0.987 (0.011) & 2.002 (0.063) & 0.993 (0.005) \\
    \hline
    \end{tabular}%
\begin{tablenotes}
    \item
    Note: Value, the value of the estimated regime in the full sample; Opt\%, the percentage of patients  in the full sample who were correctly classified to their true optimal treatments; Numbers in parentheses are empirical standard errors.
\end{tablenotes}
  \end{threeparttable}
  \label{simu_single_stage}%
\end{table}%

The results from Simulation 1 are presented in Table \ref{simu_single_stage}. When nonparametric estimation was used for the Q-function and/or treatment propensity scores, methods based on complete cases (CC) performed comparably to the All-QL (C) method applied to the 
complete data.
In particular, the performance of CC-CFBL and CC-ACFBL was the closest to   All-QL(C), demonstrating their robustness and efficiency in scenarios with missing data.
Among all CC-based methods,  CC-QL (I) and CC-EARL (II)  resulted in low proportions of patients for whom the optimal treatments were accurately identified. This is not surprising since estimators based on parametric models are vulnerable to model misspecification and thus less robust than those based on nonparametric estimation. CC-CFBL also performed better than CC-OWL (N), with this advantage becoming more pronounced with smaller sample sizes ($n=500$). This highlights the benefits of covariate balancing in finite samples over IPW with propensity scores estimated by nonparametric prediction models. Furthermore, CC-ACFBL and CC-EARL (NN) slightly outperformed CC-CFBL and CC-OWL (N), respectively, suggesting the benefit of augmenting with outcome information.

\subsection{Simulation 2. Multiple-stage scenario}

We further evaluated the performance of the proposed EE-ACFBL estimator when Assumptions 1-6 were satisfied in the multiple-stage scenario. Specifically, we simulated data from an observational study with two stages of intervention. There were two covariates $X_{t,1}$ and $X_{t,2}$ at stage $t = 1, 2$. Among these covariates, only $X_{1,2}$ and $X_{2,2}$ were partially missing, and their missingness indicators were $R_1$ and $R_2$, respectively. The detailed set-up is listed in Table \ref{simu2_DGM}.

\begin{table}[htbp]
    \centering
    \caption{Details of the  Simulation 2 set-up}
    \begin{tabular}{rcl}
    \hline
    \multicolumn{3}{l}{\textbf{Data-generating mechanism}}\\
    $\left( \begin{matrix}X_{1,1}\\X_{2,1}\end{matrix}\right)$  &  $\sim$ & $N\left(\left( \begin{matrix}0\\0\end{matrix}\right), \left(  \begin{matrix}1 & 0.5\\0.5 & 1 \end{matrix}\right) \right) $\\
    $X_{1,2}$ & $\sim$ & $\text{Uniform}(0,2)$\\
    $R_1$  & $\sim$ & $\text{Bernoulli} ([1+\exp(-3+X_{1,2})]^{-1})$\\
    $A_1$ &  $\sim$ & $2\times \text{Bernoulli}(\text{expit}(-1+2X_{1,1}^2-X_{1,2}^2-R_1)) - 1$\\
    $Y_1$ & $=$  & $-2+2A_1(1.5-X_{1,2})+2 X_{1,1}^2 + X_{1,2} + \epsilon_1 ; \epsilon_1 \sim N(0,1)$\\
    $X_{2,2}$ &$ \sim$ &  $\text{Uniform}(0,2)$ \\
    $R_2$ & $\sim$ &  $\text{Bernoulli} ([1+\exp(-1+A_1-Y_1+2X_{2,1}^2-X_{2,2})]^{-1})$ \\
    $A_2$ & $\sim$ & $2\times \text{Bernoulli} (\text{expit}(2-X_{2,1}^2+X_{2,2}-R_2) ) - 1$\\
    $Y_2$ & $=$ & $-3+A_2(1-A_1+X_{2,2}) + 2X_{1,2}- 2X_{2,1}^2 + \epsilon_2; \epsilon_2 \sim N(0,1)$\\
    \multicolumn{3}{c}{}\\
     \multicolumn{3}{l}{\textbf{Final outcome}}\\
     $Y$ & = & $Y_1+Y_2$\\
    \multicolumn{3}{c}{}\\
    \multicolumn{3}{l}{\textbf{Settings implied by the data-generating mechanism}}\\
    $Y^{pse}_1$ & = &  $-2-A_1+X_{2,2}+2X_{1,2}-2X_{2,1}^2 + Y_1$  \\
    {$d_2^{opt}(\bm{h}_2)$} & = & $2\mathbbm{I}(1-a_1+x_{2,2})-1$ \\
    {$d_1^{opt}(\bm{h}_1)$} & = & $2\mathbbm{I}(1-x_{1,2})-1$ \\
    $P(R^{pse}_1=1\mid \bm{h}_1,a_1,y^{pse}_1, R_1=1)$ & = & $[1+\exp(-1+2x_{1,2}-y^{pse}_1)]^{-1}$\\
   
    \hline
    \end{tabular}
    \label{simu2_DGM}
\end{table}

The missingness proportions for $X_{1,2}$ and $X_{2,2}$ were about $13.2\%$ and $17.3\%$, respectively. 
Since $1-a_1+x_{2,2}\geq 0$, the optimal treatment at stage 2 was always 1. The pseudo-outcome at stage 1, $Y^{pse}_1$, was equal to $-2-A_1+X_{2,2}+2X_{1,2}-2X_{2,1}^2 + Y_1$, and $Q_1(x_{1,1},x_{1,2},a_1) = -4.5+2a_1(1-x_{1,2}) +1.5x_{1,1}^2 +3x_{1,2}$. The true optimal treatment rule at stage 1 was $2\mathbbm{I}(1-x_{1,2} \geq 0)-1$. Since $R_1$ and $R_2$ were directly associated with the values of $X_{1,2}$ and
$X_{2,2}$, respectively,  $X_{1,2}$ and
$X_{2,2}$ were nonignorably missing. Besides, $R_1$ only depended on $X_{1,2}$, and $R_2$ only depended on $(A_{1}, Y_1, X_{2,1}, X_{2,2})$, hence the future-independent missingness assumption held. Moreover, in the $S_1$ sample, the missingness indicator of $Y^{pse}_1$ was the same as that for $X_{2,2}$ because $Y^{pse}_1 = -2-A_1+X_{2,2}+2X_{1,2}-2X_{2,1}^2 + Y_1$ and $(A_1, X_{1,2},X_{2,1},X_{2,2})$ were fully observed in the $S_1$ sample. Since $P(R_2=1\mid \bm{h}_2, R_1=1) = [1+\exp(-1+a_1-y_1+2x_{2,1}^2-x_{2,2})]^{-1}$, we have $P(R^{pse}_1=1\mid \bm{h}_1,a_1,y^{pse}_1) = [1+\exp(-1+2x_{1,2}-y^{pse}_1)]^{-1}$. Therefore, $Y^{pse}_1$ is nonignorably missing and $(X_{1,1},A_1)$ were conditionally independent of $R^{pse}_1$ given $(X_{1,2}, Y^{pse}_1)$. Further, $(X_{1,1},A_1)$ were directly related with $Y^{pse}_1$, thus we employ them as nonresponse instrumental variables, where the function $\Gamma_{1,\bm{\gamma}}(y^{pse}_1)= -y^{pse}_1$ in the semiparametric missingness propensity model is correctly specified.  We again generated 1000 data sets with sample sizes of $n=500$ and $n=2000$.

We compared the proposed EE-ACFBL estimator with the EE-based Q-learning method (EE-QL)  by \cite{sun2025weighted}. Specifically, we considered both the scenario that the Q-functions were correctly specified (EE-QL (C)) and the scenario that the Q-functions were misspecified (EE-QL (I)). For the EE-QL (I) method, the stage-2 and stage-1 Q-functions were misspecified as $Q(\bm{h}_2,a_2) = \beta_{20} + a_2(\psi_{20} + \phi_{2A}a_1 + \phi_{22}x_{2,2}) + \beta_{21}x_{21}$ and $Q(\bm{h}_1,a_1) = \beta_{10} + a_1(\psi_{10}+\psi_{11}x_{1,2}) + \beta_{1,2}X_{1,1}$, respectively. 
The performance of the ACFBL and Q-learning estimators with complete data and without IPW for addressing missingness  (All-ACEBL, All-QL (C), All-QL (I)) was used as benchmarks. 

\begin{table}[htbp]
  \centering
  \caption{Overall performance of the estimated optimal DTRs in  Simulation 2}
  \begin{threeparttable}
    \begin{tabular}{lcccc}
    \hline
          & \multicolumn{2}{c}{n=500}     & \multicolumn{2}{c}{n=2000} \\
          \cmidrule(lr){2-3}\cmidrule(lr){4-5}
          & Value & Opt\% & Value & Opt\% \\
          \hline
    EE-QL (I) & 1.060 (0.191) & 0.517 (0.049) & 1.011 (0.087) &  0.502 (0.016) \\
    EE-QL (C) & 1.980 (0.143) & 0.954 (0.032) & 1.993 (0.072) & 0.975 (0.018)\\
    EE-ACFBL & 1.962 (0.154) & 0.921 (0.058) & 1.979 (0.077) & 0.938 (0.042)\\
    All-QL (I) & 1.355 (0.198) & 0.603 (0.067) & 1.345 (0.108) & 0.596 (0.038)\\
    All-QL (C) & 1.987 (0.141) & 0.966 (0.024) & 1.996 (0.071) & 0.983 (0.012)\\
    All-ACFBL & 1.985 (0.143) & 0.950 (0.038) & 1.995 (0.072) & 0.968 (0.025)\\
    \hline
    \end{tabular}%
\begin{tablenotes}
    \item
    Note: Value, the value of the estimated regime in the full sample; Opt\%, the percentage of patients in the full sample who were correctly classified to their true optimal treatments in both stages; Numbers in parentheses are empirical standard errors.
\end{tablenotes}
  \end{threeparttable}
  \label{simu_multi_stage}%
\end{table}%

The values and correct classification rates of the estimated optimal DTRs are presented in Table \ref{simu_multi_stage}. 
EE-ACFBL outperformed  EE-QL (I), which was vulnerable to the misspecification of the Q-functions. 
When the parametric models for Q-functions were correctly specified,  EE-QL (C) and All-QL(C)  performed slightly better than EE-ACFBL and All-ACFBL, respectively, indicating that the ACFBL method achieved efficiency comparable to the Q-learning methods with correctly specified Q-functions.
Notably, All-QL(I) performed much worse than All-ACFBL, although it was better than EE-QL(I). This indicates that incorrect specification of the Q-function has more severe consequences on Q-learning when there are missing data issues. In contrast, EE-ACFBL performed similarly to All-ACFBL, demonstrating its robustness and efficiency in scenarios with missing data.

To further assess the impact of missingness and model misspecification on different estimation steps of DTRs, we investigated the detailed performance of the estimated DTRs by stage. Specifically, we examined the mean squared error (MSE) of $\hat{Y}^{pse}_t$ and the correct classification rates at both stages. 
For the EE-based methods, we calculated the MSE in the $S_2$ sample, while for the methods applied to the complete data, we calculated the MSE in the full sample. 
Table \ref{simu_2stage_detailed} shows that although all methods performed well in identifying the optimal treatment at stage 2, the pseudo-outcome estimate $\hat{Y}^{pse}_t$ by  EE-QL (I) and All-QL (I) exhibited large MSEs, which may lead to biased estimates of the missingness propensity and Q-function at stage 1. Consequently, while the misspecification of Q-function at stage 2  had minimal impact on stage-2 classification rates, its propagation to earlier stages through pseudo-outcomes substantially reduced the correct classification rates at stage 1. These results highlight the importance of robust pseudo-outcome estimation in mitigating error accumulation across multiple stages when handling nonignorable missing covariates.

\begin{table}[htbp]
    \centering
    \caption{Detailed performance of the estimated optimal DTRs by stages in   Simulation 2. }
  \begin{threeparttable}
    \begin{tabular}{lccc}
    \hline
     & stage 2 Opt\% & MSE of $\hat{Y}^{pse}_t$ & stage 1 Opt\% \\
    \hline
    \multicolumn{4}{c}{}\\
    \multicolumn{4}{c}{$n=500$}\\
    \hline
    EE-QL (I) & 0.999 (0.004) & 3.310 (0.581) & 0.517 (0.050) \\
    EE-QL (C) & 0.993 (0.011) & 0.016 (0.011) & 0.961 (0.030) \\
    EE-ACFBL  & 0.998 (0.006) & 0.040 (0.019) & 0.923 (0.058)\\
    All-QL (I) & 1.000 (0.003) & 3.551 (0.478) & 0.604 (0.067)\\
    All-QL (C) & 0.994 (0.010) & 0.011 (0.007) & 0.973 (0.022) \\
    All-ACFBL  & 0.999 (0.006) & 0.028 (0.013) & 0.952 (0.038)\\
    \hline
    \multicolumn{4}{c}{}\\
    \multicolumn{4}{c}{$n=2000$}\\
    \hline
    EE-QL (I) & 1.000 (0.000) & 3.294 (0.290) & 0.503 (0.016)\\
    EE-QL (C) & 0.997 (0.005) & 0.004 (0.003) & 0.978 (0.018)\\
    EE-ACFBL  & 1.000 (0.001) & 0.011 (0.005) & 0.938 (0.042)\\
    All-QL (I) & 1.000 (0.000) & 3.535 (0.245) & 0.596 (0.038)\\
    All-QL (C) & 0.997 (0.005) & 0.003 (0.002) & 0.986 (0.011)\\
    All-ACFBL  & 1.000 (0.001) & 0.008 (0.003) & 0.968 (0.025)\\
    \hline
    \end{tabular}
\begin{tablenotes}
    \item
    Note: Opt\%, the percentage of patients  in the full sample who were correctly classified to their true optimal treatments in corresponding stages. For methods applied to the complete data, the MSE is calculated using the full sample; for EE-based methods, the MSE is calculated using the $S_2$ sample. Numbers in parentheses are empirical standard errors. 
\end{tablenotes}
\end{threeparttable}
    \label{simu_2stage_detailed}
\end{table}

\subsection{Simulation 3. Scenario with mild violations of Assumption 6}

To investigate the robustness of the proposed EE-ACFBL estimator when Assumption 6 was violated, we kept all other elements of the data-generating mechanism in Table~\ref{simu2_DGM} intact, while altering the true model for $R^{pse}_1$ to $P(R^{pse}_1=1) =[1+\exp(-2-2X_{1,2}-Y^{pse}_1 + \alpha_{AX} A_1 X_{1,1} )]^{-1}$.  When $\alpha_{AX}=0$, the data-generating mechanism was the same as that in Simulation 2 and Assumption 6 was satisfied. However, when $\alpha_{AX} \neq 0$, $(A_1,X_{1,1})$ was associated  with $R^{pse}_1$ conditional on $(X_{1,2},Y^{pse}_1)$ and there were no valid nonresponse instrumental variables. We set $\alpha_{AX} \in \{-0.4,-0.2,0.2,0.4\}$ such that $A_1$ and $X_{1,1}$ were weakly associated with $P(R^{pse}_1=1)$ conditional  on $X_{1,2}$ and $Y^{pse}_1$.  We generated 1000 data sets with the sample size $n=500$.

The results of Simulation 3 are presented in Table~\ref{simu_sen}. We see that the proposed EE-ACFBL estimator was robust when $(A_1, X_{1,1})$ were weakly related to $R^{pse}_1$. Despite the slightly declining performance of the corresponding optimal DTR as the absolute value of $\alpha_{AX}$ increased, the EE-ACFBL estimator consistently outperformed the EE-QL (I) estimator, yielding much higher values and rates of correct classification.

\begin{table}[htbp]
  \centering
  \caption{Performance of the estimated optimal DTRs with mild violations of Assumption 6 in  Simulation 2 }
  \begin{threeparttable}
    \begin{tabular}{lcc}
    \hline
          & Value & Opt\%  \\
          \hline
    \multicolumn{3}{l}{ }\\
    \multicolumn{3}{c}{ $\alpha_{AX} = -0.4$ }\\
    \hline
    EE-QL (I)  & 0.574 (0.362) & 0.605 (0.079) \\
    EE-ACFBL & 1.866 (0.237) & 0.925 (0.056)  \\
    All-QL (C)  & 1.989 (0.173) & 0.965 (0.025) \\
    \hline
    \multicolumn{3}{l}{ }\\
    \multicolumn{3}{c}{ $\alpha_{AX} = -0.2$ }\\
    \hline
    EE-QL (I)  & 0.569 (0.367) & 0.604 (0.080) \\
    EE-ACFBL & 1.867 (0.239) & 0.926 (0.056)  \\
    All-QL (C) & 1.986 (0.168) & 0.965 (0.025) \\
    \hline
    \multicolumn{3}{l}{ }\\
    \multicolumn{3}{c}{ $\alpha_{AX} = 0.2$ }\\
    \hline
    EE-QL (I)  & 0.578 (0.363) & 0.606 (0.079) \\
    EE-ACFBL & 1.865 (0.242) & 0.924 (0.056)  \\
    All-QL (C) & 1.986 (0.174) & 0.964 (0.025) \\
    \hline
    \multicolumn{3}{l}{ }\\
    \multicolumn{3}{c}{ $\alpha_{AX} = 0.4$ }\\
    \hline
    EE-QL (I)  & 0.595 (0.357) & 0.610 (0.077) \\
    EE-ACFBL & 1.854 (0.247) & 0.920 (0.058)  \\
    All-QL (C) & 1.985 (0.175) & 0.963 (0.026) \\
    \hline
    \end{tabular}%
\begin{tablenotes}
    \item
    Note: Value, the value of the estimated regime in the full sample; Opt\%, the percentage of patients in the full sample who were correctly classified to their true optimal treatments in both stages; Numbers in parentheses are empirical standard errors.
\end{tablenotes}
  \end{threeparttable}
  \label{simu_sen}%
\end{table}%

\section{Application to the MIMIC-III data}\label{application}

We applied the proposed methods to the MIMIC-III data \citep{Johnson_et_al_2016} for investigating the optimal 2-stage fluid strategy to treat sepsis patients in ICUs. 
We followed the same patient selection criteria as outlined by \cite{sun2025weighted}, focusing on adult septic patients admitted to the medical ICU after initially presenting to the emergency department. The detailed cohort eligibility criteria can be found in Figure S1 in the Supplementary Materials. 
Following \cite{Speth_et_al_2022} and \cite{sun2025weighted}, the stage 1 and stage 2 treatments were categorized as either fluid restrictive ($<$30 ml/kg) or fluid liberal ($\geq$ 30 ml/kg) strategies in 0-3 hours and 3-24 hours post-ICU admission, respectively. The baseline covariates at stage 1 included gender, age, weight, racial groups, and Elixhauser comorbidity score. For the variables considered before stage 2 treatment, in addition to the above baseline covariates, we also included the following intermediate variables: use of mechanical ventilation and vasopressors within the first 3-hour period, the patient's Sequential Organ Failure Assessment (SOFA)  score evaluated at 3 hours post-admission and hemodynamic variables including heart rate, blood pressure, saturation of peripheral oxygen (SpO2), respiratory rate, temperature at 3 hours post-admission and urine output within 0-3 hours post-admission. These hemodynamic variables contained missing values, which were likely to be nonignorable due to the informative monitoring of the patients in the MIMIC-III database, as discussed in Section~\ref{motivatingdata}. The missing proportions of the hemodynamic variables are provided in Table~S2 of the Supplementary Materials.

The outcome of interest was the SOFA score evaluated at 24 hours post-admission. The SOFA score is a clinical tool used to assess the severity of organ dysfunction by assigning scores to various organ systems. A higher SOFA score indicates a more severe impairment. In our analysis, we used the negative value of the SOFA score at 24 hours post-admission as the final outcome to be maximized by the optimal fluid resuscitation strategy.

There were 973 patients in the selected cohort, among them $53.2\%$ were male and $78.7\%$ were Caucasian. 45.5\% of the patients received the fluid liberal strategy in 0-3 hours post-admission while $53.5\%$ received the fluid liberal strategy in 3-24 hours post-admission. Overall, 67.1\% of patients had fully observed covariates at stage 2, while the rest of the patients had at least one covariate with missing values.

As discussed in Section~\ref{CCsame}, it was plausible that Assumption 4 holds; that is, the missing data mechanism of the hemodynamic variables at 3 hours post-admission was unrelated to the patient’s SOFA score at 24 hours post-admission, conditioning on the patient's medical history and treatments up to stage 2. 
Denote the negative of SOFA score at 24 hours post-admission under the optimal stage 2 treatment as $Y^{pse}_1$. It was plausible that weight and gender were independent of the missingness of hemodynamic variables given other baseline covariates and $Y^{pse}_1$ since informative monitoring of ICU patients was likely to be associated with the severity of sepsis disease progression (reflected in $Y^{pse}_1$) and not directly depend on baseline characteristics of the patients. 
Following \cite{sun2025weighted}, we employed weight, gender, and Elixhauser comorbidity score as candidates for the nonresponse instrumental variables and investigated the impact of using different sets of candidates for nonresponse instrumental variables on the results obtained by the EE-ACFBL estimator. Specifically, we used all the covariates in stage 1 except the nonresponse instrumental variables as $\bm{U}_1$ in model~\eqref{missingkernel}, and 
let $\Gamma_{1,\gamma_1}(y_1^{pse})=\gamma_1y^{pse}_1$. 
For comparison, we also applied the EE-based Q-learning (EE-QL) method by \cite{sun2025weighted} with linear models and the same nonresponse instrumental variables. Furthermore, the ACFBL estimator and Q-learning without the hemodynamic variables were implemented to assess the benefit of incorporating the hemodynamic variables in obtaining optimal fluid resuscitation strategies. Cross-validation was conducted to assess the performance of the estimated optimal DTRs. More details of the estimation procedure are provided in Appendix H of the Supplementary Materials.

\begin{table}[htbp]
\caption{Improvement of SOFA at 24 hours post-admission for sepsis patients in the selected cohort of  the MIMIC-III database under the estimated optimal two-stage fluid strategies using the EE-QL and EE-ACFBL estimators with different nonresponse instrumental variables as well as the QL and ACFBL estimators ignoring the hemodynamic variables.}\label{RDAtable1}
  \centering
  \begin{threeparttable}
    \begin{tabular}{lll}
    \hline
     & Mean & Std  \\
    \hline 
    EE-QL (ECS)    &  1.297 & 0.163  \\
    EE-QL (weight)   & 1.432 & 0.156  \\
    EE-QL (ECS, weight)   & 1.338 & 0.157\\
    EE-QL (gender)   & 1.430 & 0.181 \\
    EE-QL (gender, ECS)   & 1.338 & 0.167 \\
    EE-QL (gender, weight)    & 1.459 & 0.165\\
    EE-QL (gender, weight, ECS)    & 1.345 & 0.170 \\
    EE-ACFBL (ECS)    &  1.315 & 0.164  \\
    EE-ACFBL (weight)   & 1.430 & 0.161  \\
    EE-ACFBL (ECS, weight)   & 1.344 & 0.165\\
    EE-ACFBL (gender)   & 1.435 & 0.185 \\
    EE-ACFBL (gender, ECS)   & 1.341 & 0.170 \\
    EE-ACFBL (gender, weight)    & 1.463 & 0.171\\
    EE-ACFBL (gender, weight, ECS)    & 1.343  & 0.179\\
    NH-QL &  0.909 & 0.102 \\
    NH-ACFBL &   1.053 &  0.112  \\
    \hline
    \end{tabular}%
\begin{tablenotes}
    \item Note: 
    EE-, methods employing  estimating equations to obtain the missingness propensity of the pseudo-outcome involving hemodynamic variables;
    NH-, methods did not include hemodynamic variables in their models. 
    The variables in parentheses are nonresponse instrumental variables.
    ECS, Elixhauser comorbidity score; 
    Mean, the average of estimated improvement  of SOFA based on cross-validation using 1000 random splits; 
    Std, the standard error of estimated improvement  of SOFA based on cross-validation using 1000 random splits. 
\end{tablenotes}
    \end{threeparttable}
\end{table}%

Table \ref{RDAtable1} presents the averages and standard errors of estimated improvement of SOFA based on cross-validation using 1000 random splits.
EE-ACFBL and EE-QL had similar means and standard errors for the estimated improvement of SOFA score under the estimated regime relative to the observed SOFA score at 24 hours post-admission, suggesting that the Q-function estimates in EE-QL were approximately equivalent to the smoothing spline estimates from EE-ACFBL. However, while both methods yielded similar results for a given set of the three candidate nonresponse instrumental variables, including the Elixhauser comorbidity score in the set of nonresponse instrumental variables led to a reduced estimated improvement of SOFA at 24 hours post-admission. This may be because informative monitoring of ICU patients (thus the missingness of hemodynamic variables) was also associated with patients' comorbidities, even after adjusting for sepsis disease progression. As a result, Elixhauser comorbidity score was not a valid nonresponse instrumental variable. Moreover, EE-ACFBL and EE-QL both produced considerably higher means of the estimated improvement of the final outcome than their counterparts ignoring the hemodynamic variables with missing data.   Therefore, our analysis results confirmed the conclusion in \cite{sun2025weighted} that it is important to consider hemodynamic variables and handle their nonignorable missingness when formulating fluid resuscitation strategies for sepsis patients during the 3-24 hours following admission to medical ICUs. 

\section{Discussion}

In this article, we proposed robust estimators of optimal single-stage and multi-stage treatment rules within the direct-search framework for challenging scenarios with nonignorable missing covariates. By constructing a robust and efficient value estimator with RKHS-based nonparametric methods, we mitigated the misspecification risk induced by parametric Q-functions and further amplified by the biased estimates of the missingness propensity of the pseudo-outcomes in the weighted Q-learning approach by \cite{sun2025weighted}, as shown in Simulations 2 and 3. Moreover, replacing treatment propensity weights estimated by prediction models with more stable balancing weights also resulted in better performance of the proposed estimators than other competitors in Simulation 1 for single-stage scenarios. In the MIMIC-III data analysis,  we confirmed the findings in \cite{sun2025weighted} that incorporating hemodynamic variables into optimal fluid strategy estimation and handling nonignorable missing covariates can improve the short-term outcome of sepsis patients in ICUs.

There are several directions for future work. First, although we focus on the kernel-based methods with the assumption that the outcome mean models belongs to a RKHS in this article, other nonparametric models can also be employed for the estimation of covariate balancing weights \citep{chan2016globally} and outcome models when the assumptions required by smoothing splines do not hold. Second, the dimension of the covariates is often high in practice, especially in multiple-stage scenarios. In this case, the kernel regression method used in the missingness propensity model would face the `curse of dimensionality’ problem. To address this, we may consider dimension reduction techniques proposed by \cite{Tang_Zhao_Zhu_2014}. Last, our methods utilize inverse probability weighting in the sample with complete covariates at different stages, whose size can be small when the majority of patients have at least one covariate with missing values. In this case, we could explore the multiple imputation method under nonignorable missingness \citep{tompsett2018use}.

\bibliographystyle{biorefs}
\bibliography{robust_DTR_MNAR_bib}


\end{document}